\documentclass[12pt]{iopart}

\usepackage{graphicx}
\usepackage{siunitx}
\usepackage[colorlinks=true, linkcolor=blue,citecolor=red]{hyperref}
\usepackage{xcolor}
\usepackage{ulem}
\newcommand{\R}{\mathcal{R}}
\newcommand{\Rm}{\mathcal{R}_{\rm m}}

\newcommand{\kmo}{k_m^{(0)}}

\newcommand{\ii}{{\rm i}}
\newcommand{\ee}{{\rm e}}
\newcommand{\new}[1]{\textcolor{black}{#1}}

\newcommand{\be}{\begin{equaion}}
\newcommand{\bea}{\begin{eqnarray}}
\newcommand{\eea}{\end{eqnarray}}

\definecolor{green}{rgb}{0.1,.7,0.05}

\begin{document}

\title{Two--membrane cavity optomechanics}

\author{Paolo~Piergentili$^{1,2}$, Letizia~Catalini$^{1}$, Mateusz~Bawaj$^{1,2}$, Stefano~Zippilli$^{1,2}$, Nicola~Malossi$^{1,2}$, Riccardo~Natali$^{1,2}$, David~Vitali$^{1,2,3}$, Giovanni~Di~Giuseppe$^{1,2}$}
\address{	$^1$ \text{School of Science and Technology, Physics Division, University of Camerino, }\\
		\quad\text{I-62032 Camerino (MC), Italy}\\
		$^2$ \text{INFN, Sezione di Perugia, Italy}\\
		$^3$ \text{CNR-INO, L.go Enrico Fermi 6, I-50125 Firenze, Italy}}

%
%\author{Content \& Services Team}
%
%\address{IOP Publishing, Temple Circus, Temple Way, Bristol BS1 6HG, UK}
\ead{david.vitali@unicam.it, gianni.digiuseppe@unicam.it}

\vspace{10pt}
\begin{indented}
\item[\today]
\end{indented}

\begin{abstract}
We study the optomechanical behaviour of a driven Fabry--P\'erot cavity containing two vibrating dielectric membranes. We characterize the cavity--mode frequency shift as a function of the two-membrane positions, and report a $\sim 2.47$ gain in the optomechanical coupling strength of the membrane relative motion with respect to the single membrane case. This is achieved when the two membranes are properly positioned to form an inner cavity which is resonant  with the driving field.
We also show that this two-membrane system has the capability to tune the single-photon
optomechanical coupling on demand, and represents a promising platform for implementing cavity optomechanics with distinct oscillators.
Such a configuration has the potential to enable cavity optomechanics in the strong single-photon coupling regime,
and to study synchronization in optically linked mechanical resonators.
\end{abstract}

\date{\today}% It is always \today, today,

%\tableofcontents

\section{Introduction}
Multi--element systems of micro/nano--mechanical resonators offer promising prospects for enhanced optomechanical performances
~\cite{Bhattacharya:2008ab,Hartmann:2008aa,Xuereb:2012fk,Xuereb:2013ys,Li:2016aa,Nair:2016aa,Li:2017ac}, coherent control~\cite{Ludwig:2013aa,Weaver:2017aa},
and for the exploration of multi--oscillators synchronization~\cite{Heinrich:2011ab,Holmes:2012aa,Zhang:2012aa,Ludwig:2013aa,Bagheri:2013ht,Agrawal:2013fk,Matheny:2014fv,Zhang:2015ad}.
The standard path for reaching the strong single-photon optomechanical coupling regime is to consider co--localized optical and vibrational modes~\cite{Baker:2014aa,Balram:2014aa,Meenehan:2014aa}, with a large spatial overlap confined in very small volumes, corresponding to mechanical modes with extremely small effective mass.
An alternative solution, capable of providing systems with orders of magnitude increased ratio between the single-photon optomechanical coupling rate, and the cavity decay rate, is to exploit quantum interference in multi--element optomechanical setups~\cite{Xuereb:2012fk,Xuereb:2013ys,Li:2016aa}.
Although the simplest two-membrane sandwich in an optical cavity is a paradigm for the realization of strong-coupling optomechanics, and the observation of collective mechanical effects (such as synchronization), no experimental studies of these phenomena have been reported till now. Previous related results~\cite{Nair:2017ab} were confined
only to the optical and mechanical characterization of two--membrane sandwiches.

Here we report on the first experimental characterization of the optical, mechanical, and especially optomechanical properties of a sandwich constituted of two parallel membranes within an optical cavity. We show how the
resonance frequencies of the optical cavity are shifted as a function of the position of the two membranes. This effect is central to the description of the optomechanical properties of the system, since it provides a direct estimation of the strength of the
couplings~\cite{Bhattacharya:2008ab,Bhattacharya:2008aa,Jayich:2008nx,Thompson:2008}.
By investigating the shifts of the cavity resonances we find that the optomechanical coupling strength is enhanced by constructive interference when the two membranes are positioned to form an inner cavity which is resonant with the driving field.
Specifically we  determine a gain of $\sim 2.47$ in the coupling strength of the relative mechanical motion with respect to the single membrane configuration.  We finally prove both the capability to tune on demand the single-photon optomechanical couplings, and the simultaneous optical cooling of the fundamental modes of the two distinct membranes.
%

%------------------------------------------------------------------
\section{Theory}
\label{sec:theory}

Generalizing the results obtained in Ref.~\cite{Li:2016aa}, we consider the case of two different movable dielectric membranes placed inside a Fabry--P\'erot cavity of length $L$, which is driven by an external laser. The Fabry--P\'erot cavity is composed of two identical mirrors with electric field reflection and transmission coefficients $r$ and $t$, respectively.
The membranes can be modelled as dielectric slabs of thickness $L_{{\rm m},j}$ and index of refraction $n_j$ (where the index $j=1,2$ distinguish the parameters of the two membranes), such that their reflection and transmission coefficient can be expressed as
\begin{eqnarray}
    r_j = \frac{(n_j^2-1)\sin(kn_jL_{{\rm m},j})}
    				{(n_j^2+1)\sin(kn_jL_{{\rm m},j}) +  2\ii\, n_j\cos(kn_jL_{{\rm m},j})} 	
	\label{eq:rm}\\
    t_j = \frac{2n_j}
    				{(n_j^2+1)\sin(kn_jL_{{\rm m},j}) +  2\ii\, n_j\cos(kn_jL_{{\rm m},j})}		
	\label{eq:tm}		
\end{eqnarray}
where $k = 2\pi/\lambda$ is the wavenumber of the electric field, and $\lambda$ is its wavelength.

\begin{figure}[!b]
\centering
\includegraphics[width=0.45\linewidth]{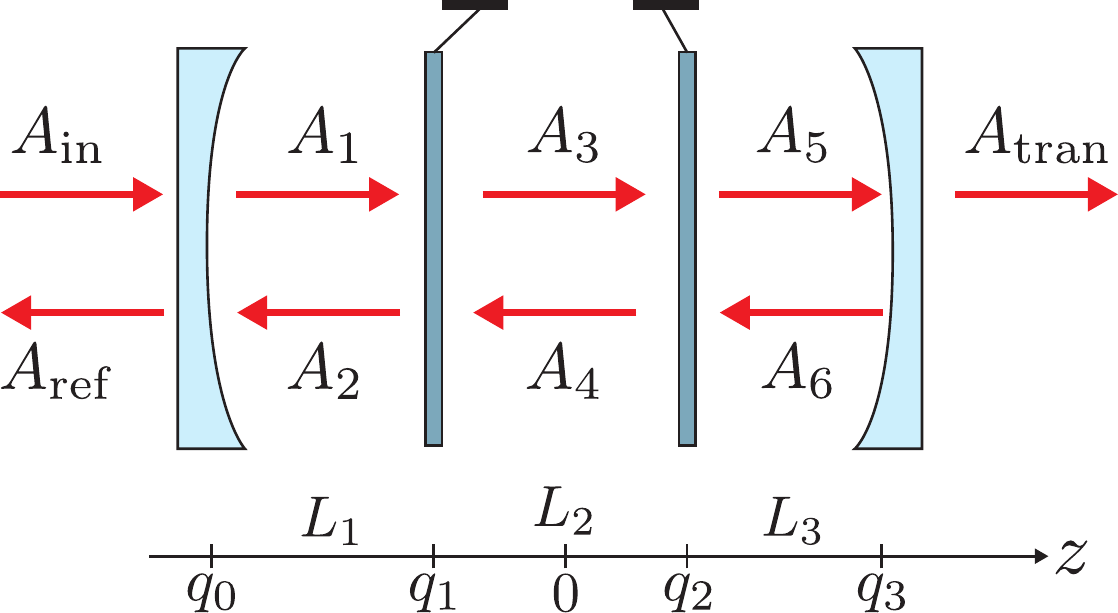}
\caption{Schematic diagram of the system. Two movable dielectric membranes are placed inside a Fabry--P\'erot cavity of length $L$ which is driven by an external laser. The position of two fixed mirrors (movable membranes) is denoted by $q_0$ and $q_3$ ($q_1$ and $q_2$); we have $L_i=q_i-q_{i-1}$ ($i=1,2,3$), with
$q_3=-q_0=L/2$. }
\label{fig:Figure_1}
\end{figure}
The optical resonance frequencies correspond to the maxima of transmission of the whole cavity. The electric field amplitudes $A_j$ of incident ($j=\text{in}$), reflected ($j=\text{ref}$), and transmitted ($j=\text{tran}$) waves, as well as for the fields in the cavity ($j=1,2,\dots,6$) (see Figure~\ref{fig:Figure_1}), satisfy the following equations:
\begin{eqnarray}
A_1&=\ii\,t\,A_\text{in} - r\,A_2 e^{\ii\,k L_1},  \label{Aeqs3}\\
A_2&=\ii\,t_1 A_4 e^{\ii\,k L_2} - r_1 A_1 e^{\ii\,k L_1},  \\
A_3&=\ii\,t_1 A_1 e^{\ii\,k L_1} - r_1 A_4 e^{\ii\,k L_2},  \\
A_4&=\ii\,t_2 A_6 e^{\ii\,k L_3} - r_2 A_3 e^{\ii\,k L_2},  \\
A_5&=\ii\,t_2 A_3 e^{\ii\,k L_2} - r_2 A_6 e^{\ii\,k L_3},  \\
A_6&=-r\,A_5 e^{\ii\,k L_3},  \\
A_\text{ref}&=\ii\,t\,A_2 e^{\ii\,k L_1} - r\,A_\text{in},  \\
A_\text{tran}&=\ii\,t\,A_5 e^{\ii\,k L_3},
\label{Aeqs}
\end{eqnarray}
where
$L_j=q_j-q_{j-1}$ (with $q_j$ the positions of the various elements defined in Figure~\ref{fig:Figure_1}, and $j$=1,2,3)
is the length of the subcavities formed by the mirrors and the membranes,
so that $L=L_1+L_2+L_3$. We point the reader to Ref.~\cite{Jayich:2008nx} for a similar approach in the case of a single membrane. Here we use the same convention of Ref.~\cite{Jayich:2008nx}
for the scattering matrix of a single scattering element, either the cavity mirror or the membrane. This is a bit different from the choice of Ref.~\cite{Li:2016aa}, which is reproduced by replacing $r$ with $-r$ into the equations above.
Eqs.~(\ref{eq:rm})--(\ref{Aeqs}) are valid, for any value of the thickness,
in the ideal one-dimensional case of plane waves, and flat,
 aligned mirrors and membranes.
They can be applied also to the case of Gaussian cavity modes and spherical external mirrors as long as the membranes are placed within the Rayleigh range of the cavity.

The system of Eqs.~(\ref{Aeqs3})--(\ref{Aeqs}) can be solved to determine the transmission coefficient
of the whole cavity. It is given by
\begin{equation}
\frac{A_\text{tran}}{A_\text{in}}=\tau_c=\frac{t^2 t_1 t_2 e^{ \ii\,  k L }}{{\cal D}},
\end{equation}
with
\begin{eqnarray}
	{\cal D} = 1 &-& r^2 (t_1^2+r_1^2)(t_2^2+r_2^2) \ee^{2  \ii\, k L} +r r_2 (t_1^2+r_1^2) \ee^{2  \ii\, k (L_1+L_2)}
				\nonumber \\	
		&+&r^2 r_1 r_2 \ee^{2  \ii\, k (L_1+L_3)}+r r_1 (t_2^2+r_2^2) \ee^{2  \ii\, k (L_2+L_3)}
				\nonumber \\
		&-&r r_1  \ee^{2  \ii\, k L_1}-r_1 r_2  \ee^{2  \ii\, k L_2}-r r_2  \ee^{2  \ii\, k L_3}.
\label{denom}
\end{eqnarray}
This last expression reproduces Eq. (4) of Ref.~\cite{Li:2016aa} when $r_1=r_2$, $t_1=t_2$, and we restrict to the case of real $n_j$, implying in particular ${\rm arg}(r_j)={\rm arg}(t_j)\equiv\phi_j$
so that $r_j^2+t_j^2=e^{2 \ii  \phi_j}$. Moreover it reproduces also the case of a single membrane which is obtained by taking $r_2 = 0$, $t_2 = - \ii  $, $L_3 =0$.
From Eqs.~(\ref{Aeqs3})--(\ref{Aeqs}) one can also derive the expression for the reflectivity, given by
\begin{eqnarray}
\frac{A_\text{refl}}{A_\text{in}} = \varrho_c=
	-r + \frac{t^2 r_1 e^{2 \ii\, k L_1 }}{1-r r_1 e^{2 \ii\, k L_1 }}
	-\frac{t^2 t_1^2 e^{2 \ii\, k (L_1+L_2) }\left[r_2-r(r_2^2+t_2^2)e^{2 \ii\, k L_3}\right]}
			{(1-r r_1 e^{2 \ii\, k L_1 }){\cal D}}\,.			
\end{eqnarray}
An explicit equation for the cavity mode frequencies can be found
in the case of negligible optical absorption of the membranes, i.e. for real $n_j$. In this case we
rewrite $r_j$ with $j=1,2$ in terms of the intensity reflectivity $R_j$ as
$r_j=\sqrt{R_j}e^{\ii\, \phi_j}$, and we assume for simplicity $r$ and $t$ real
so that we express them in terms of the corresponding intensity reflectivitys as
$r=-\sqrt{R}$, $t=-\sqrt{1-R}$. Accordingly, Eq. (\ref{denom}) becomes
\begin{eqnarray}
	{\cal D} =1 &-& R \ee^{2  \ii\, k L+2 \ii\, \phi_1 +2 \ii\, \phi_2}-\sqrt{R R_2}\ee^{2  \ii\, k (L_1+L_2)+2 \ii\, \phi_1 + \ii\, \phi_2} \label{denom2} \nonumber \\
	&+&R \sqrt{R_1 R_2} \ee^{2  \ii\, k (L_1+L_3)+ \ii\, \phi_1 + \ii\, \phi_2}
		-\sqrt{R R_1} \ee^{2  \ii\, k (L_2+L_3)+ \ii\, \phi_1 +2 \ii\, \phi_2}
				\nonumber \\
	&+&\sqrt{R R_1}  \ee^{2  \ii\, k L_1+ \ii\, \phi_1}
		-\sqrt{R_1 R_2}  \ee^{2  \ii\, k L_2+ \ii\, \phi_1 + \ii\, \phi_2}
				\nonumber \\
	&+& \sqrt{R R_2}  \ee^{2  \ii\, k L_3+ \ii\, \phi_2}.
\end{eqnarray}
The cavity mode frequencies correspond in general to the maxima of the transmission, and therefore the minima of $ |{\cal D}|^2$. In the limiting case of perfect external mirrors, $R=1$, these maxima become poles of the transmission and the modes correspond to the zeros of ${\cal D}$. In order to get a simple expression for the poles we restrict to this limiting situation which, as we have seen in Ref.~\cite{Li:2016aa}, works also
in the case of realistic high-finesse cavities for which typically $1-R\sim10^{-5}$.
In particular using the definitions $L_1=q_1+L/2$,  $L_3=L/2-q_2$ and introducing the relative coordinate $q=L_2=q_2-q_1$ we find that Eq.~(\ref{denom2}) can be rewritten, for $R=1$, as
\begin{eqnarray}
	-{\cal D}/2 \ii  =&& \sin (k L+\phi_1 +\phi_2)-\sqrt{R_1 R_2}  \sin(k L- 2 k q)\label{denom4} \nonumber \\
				&&\quad-\sqrt{ R_1} \sin(2k q_1- \phi_2)+\sqrt{ R_2} \sin(2k q_2+ \phi_1)\,.
\end{eqnarray}
By setting this  equation equal to zero we get the implicit equation for the cavity mode frequencies valid in the limit of $R \sim 1$ and for the general case of two different membranes. It reproduces the implicit equation in the two special cases of equal membranes and of one membrane only. Specifically, in the case of equal membranes $R_1=R_2=R_m$, $\phi_1=\phi_2=\phi$, and using the definitions $L'\equiv L+2\phi/k$ and $q'\equiv q+\phi/k$, $q_1=Q-q/2$, $
q_2=Q+q/2$, where $Q=(q_1+q_2)/2$ is the center--of--mass (CoM) coordinate, we get
\begin{equation}
\sin (k L')-R_m  \sin(k L'- 2 k q')+2\sqrt{ R_m} \cos(2kQ)\sin(k q')=0,
\end{equation}
which coincides with Eq. (8) of Ref.~\cite{Li:2016aa}. Instead in the one membrane case, putting $R_2=0$, $\phi_1= \phi$ and $\phi_2 = -\pi/2$, we get
\begin{eqnarray}
	-\cos (k L+\phi)-\sqrt{ R_1} \cos(2k q)=0,\nonumber
\end{eqnarray}
which is just the corresponding equation used in Ref.~\cite{Serra:2016aa} in the limit $R=1$.

In the general case the implicit equations for the mode frequencies ${\cal D}=0$, with $\cal D$ given in
Eq.~(\ref{denom4}), can be expressed using the definitions
$L'\equiv L+\phi_1/k+\phi_2/k$, $q'\equiv q+\phi_1/2k+\phi_2/2k$, and $Q'\equiv Q+\Delta\phi/4k$, ($\Delta\phi=\phi_1-\phi_2$), as
\begin{equation}
{\cal A}(kq')\sin(kL')+{\cal B}(kq')\cos(kL')={\cal F}(kQ',kq'),
\label{ABF}
\end{equation}
with ${\cal A}(kq')=1-\sqrt{R_1 R_2} \cos(2kq')$, ${\cal B}(kq')=\sqrt{R_1 R_2} \sin(2kq')$, and ${\cal F}(kQ',kq')=\sqrt{R_1}\sin(2kQ'-kq')-\sqrt{R_2}\sin(2kQ'+kq')$.
This can be further simplified with the definitions
${\tilde O}=O/\sqrt{{\cal A}^2+{\cal B}^2}$, $O={\cal A}, {\cal B}, {\cal F}$ such that
Eq.~(\ref{ABF}) can be rewritten in the equivalent form
\begin{equation}
\sin\left[kL'+\theta(kq')\right]=\tilde{\cal F}(kQ',kq'),
\label{kequation}
\end{equation}
where
\begin{equation}\label{efk}
    \tilde{\cal F}(kQ',kq') = \frac{\sqrt{R_1}\!\sin(2kQ'\!\!-\!kq')\!-\!\sqrt{R_2}\!\sin(2kQ'\!\!+\!kq')}
    					    {\sqrt{1+R_1 R_2-2\sqrt{R_1 R_2} \cos(2kq')}},
\end{equation}
and
\begin{eqnarray}\label{tetk}
	\theta(kq') = \arcsin[\tilde{\cal B}(kq')]
%				 \\
			=\arcsin\left\{\frac{\sqrt{R_1R_2}\sin(2kq')}{\sqrt{1+R_1 R_2-2\sqrt{R_1 R_2} \cos(2kq')}}\right\}\,,
\end{eqnarray}
and which, in turn, is equivalent to its formal solution obtained by inverting the $\sin$ function, that, using the definition of $L'$, can be expressed as
\begin{equation}
	k\,L=\ell\,\pi+\pi\,{\cal H}(kQ',kq')
	\label{keq2}
\end{equation}
with
\begin{equation}\label{H}
	\pi\,{\cal H}(kQ',kq')=
		(-1)^\ell \arcsin[\tilde{\cal F}(kQ',kq')]
			-\theta(kq')-\phi_1-\phi_2\ .
\end{equation}
and $\ell$ integer.
For each value of $\ell$ one finds a solution for a cavity mode wave--number $k_\ell$ that can be decomposed as the sum, $k_\ell=k_\ell^{(0)}+\delta k_\ell$,
of the empty cavity solution $k_\ell^{(0)}=\ell\,\pi/L$ (which corresponds to the condition
$R_\text{1}=R_\text{2}=0$, that implies $\tilde{\cal F}(kQ',kq')=\theta(kq')=0$) and the shift due to the membranes that is given by the implicit expression $\delta k_\ell=L^{-1}\ \pi\,{\cal H}(k_\ell Q',k_\ell q')$. In
typical experiments, $\lambda=2\pi/k_\ell^{(0)} \ll L$, so that
$\ell$ is a very large integer and this implies $ k_{\ell}^{(0)} \gg \delta k_{\ell}$. In this limit
one can safely take $L'\simeq L+\phi_1/k_{\ell}^{(0)}+\phi_2/k_{\ell}^{(0)}$ and $q'\simeq q+\phi_1/2k_{\ell}^{(0)}+\phi_2/2k_{\ell}^{(0)}$.
Correspondingly, for $R_1$ and $R_2$ not too close to one, and for not too large values of $q_1$ and $q_2$, i.e., when $q_1/L, q_2/L\ll 1$, (see Ref.~\cite{Li:2016aa}), one can safely
express the shift explicitly as a function of the empty cavity solution as $\delta k_\ell=L^{-1}\ \pi\,{\cal H}(k_\ell^{(0)}Q',k_\ell^{(0)}q')$, that can be also written
as an equation for the cavity mode frequency shift~\cite{Li:2016aa}
\begin{equation}\label{eq:approxzero}
	\delta \omega \equiv c\,\delta \kmo = \frac{\pi c}{L}\mathcal{H}\left(2\pi \frac{q_1}{\lambda},2\pi \frac{q_2}{\lambda}\right)\,.
\end{equation}
\begin{figure}[!t]
\centering
\includegraphics[width=.5\linewidth]{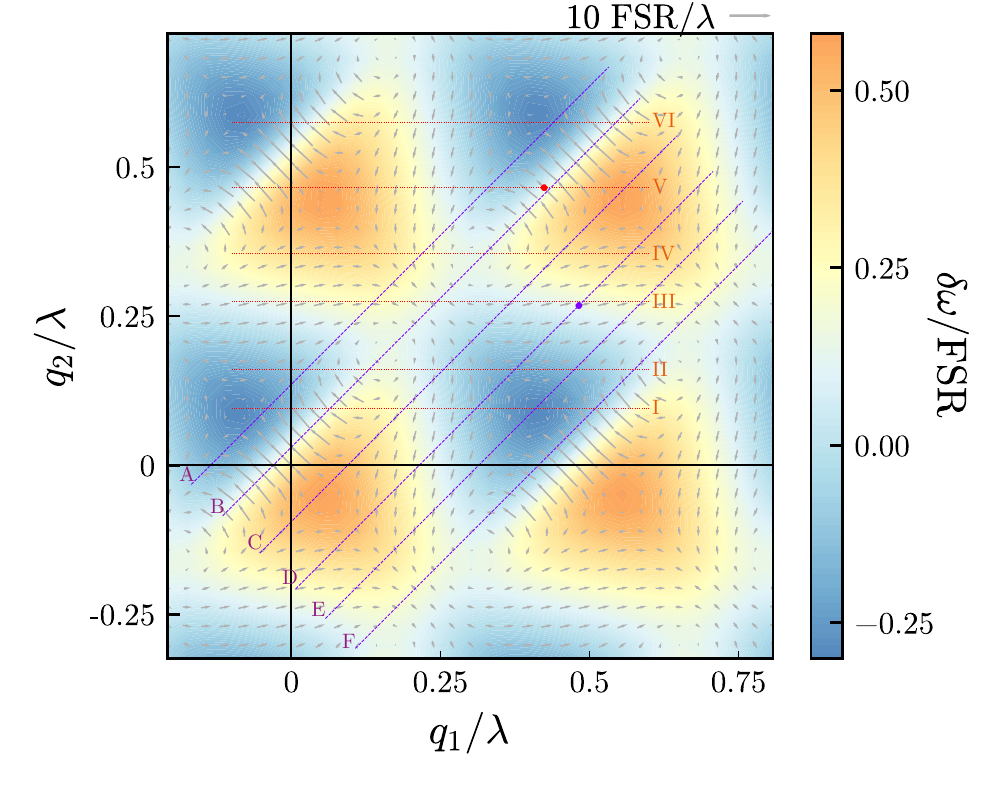}
\caption{Contour plot of the frequency shift function $\delta\omega = c\, \delta \kmo $ for even $m$ normalized to the free--spectral--range of the cavity, ${\rm FSR} = \pi c/L$, as a function of the membrane positions $q_1$ and $q_2$ normalized to the wavelength, due to the presence of the two--membrane cavity. The parameters used for the numerical analysis are: $\lambda = \SI{1064}{\nano\meter}$, $\R = \num{0.99994}$, $L = \SI{90}{\milli\meter}$, $L_{\rm m} = \SI{104}{\nano\meter}$, and $n = 2.17$.
Superimposed the vector plot of the gradient field of the frequency shift, whose components give the two optomechanical couplings, with the unit indicated on the top--right of the panel. The oblique blue lines (A--F) indicate the experimental spectra obtained by varying the CoM of the membrane--cavity system for different positions $q_2$, and reported in Figure~\ref{fig:Figure_6}. The horizontal red lines (I--VI) indicate the experimental spectra obtained by varying $q_1$ for different positions $q_2$, and reported in Figure~\ref{fig:Figure_7}. The red and blue dots represent the points where the optomechanical coupling was estimated.}
\label{fig:Figure_2}
\end{figure}
\new{This treatment in the general case of two different membranes generalizes previous results and has the advantage of providing a unique framework in which one can immediately compare the single and two-membrane case. On the other hand, for a given value of the maximum available membrane reflectivity $R_{\rm max}={\rm max}\{R_1,R_2\}$, we have numerically verified that the largest optomechanical couplings are achieved when the two membranes have identical reflectivities. For this reason we have focused our experiments to the case of nominally identical membranes, and we shall restrict from now on to this latter case.
In particular, introducing} the parameters $\Rm =R_1 = R_2$, and $\phi=\phi_1 = \phi_2$, $L_{\rm m}=L_{\rm m,1}=L_{\rm m,2}$ and $n=n_1=n_2$, the explicit dependence upon the variables $kq_1$ and $kq_2$
of the parameters $\tilde{\mathcal{F}}(kq_1,kq_2)$ and $\theta(kq_1,kq_2)$ that enter into the definition of ${\cal H}$ in Eq.~(\ref{H}), is easily obtained from Eqs. (\ref{efk}) and (\ref{tetk}), so that for identical membranes one has
\begin{eqnarray}
	\tilde{\mathcal{F}}(kq_1,kq_2) =&  -\frac{2\sqrt{\Rm}\cos[k(q_1 + q_2)]\sin[k(q_2 - q_1) + \phi]}
			{\sqrt{1 + \Rm^2 - 2\Rm\cos[2k(q_2 - q_1) + 2\phi]}},	\\
	\theta(kq_1,kq_2) =& \arcsin\left[\frac{\Rm\sin[2k(q_2 - q_1) + 2\phi]}
			{\sqrt{1 + \Rm^2 - 2\Rm\cos[2k(q_2 - q_1) + 2\phi]}}\right]\,.
\end{eqnarray}
Figure~\ref{fig:Figure_2} shows the mode frequency shift $\delta\omega$ normalized to the free--spectral--range of the cavity, ${\rm FSR} = \pi c/L$, as a function of the membrane positions $q_1$ and $q_2$ normalized to the wavelength, assuming the parameters of the experimental setup, i.e., $\lambda = \SI{1064}{\nano\meter}$, $\R = \num{0.99994}$, $L = \SI{90}{\milli\meter}$, $L_{\rm m} = \SI{104}{\nano\meter}$, and $n = 2.17$.
It is worth noting that a nonzero value of the phase $\phi$ determines a displacement of the pattern along the bisector of the second and fourth quadrants, and a constant shift of the cavity frequencies.

The optomechanical couplings strength $G_j$ are the derivative of the optical mode frequencies with respect to the position of the $j$-th membrane $q_j$.
Defining the scaled dimensionless positions $\tilde{q}_j = q_j/\lambda$, we can write in general
\begin{equation}
	G_j  = \frac{FSR}{\lambda}\,
			\frac{\partial \mathcal{H}\left(2\pi \tilde{q}_1,2\pi \tilde{q}_2\right)}{\partial \tilde{q}_j}\,,
	\label{eq:G}
\end{equation}
%
%where $x_{j}^{\rm zpf} = \sqrt{\hbar/2m_j\omega_{\rm m}^{(j)}}$ is the zero point position fluctuations of the $j$-th mechanical mode, and $\Theta_j$ is the dimensionless transverse overlap between the $j$-th mechanical mode and the optical cavity mode.

In the case of a single membrane the single-photon optomechanical coupling has the same structure of Eq.~(\ref{eq:G})
\begin{equation}
	G_{\rm sing} =  \frac{FSR}{\lambda}\,
			\frac{\partial \mathcal{H}_{\rm sing}\left(2\pi \tilde{q}\right)}{\partial \tilde{q}}\,,
	\label{eq:Gsing}
\end{equation}
but with a different dimensionless frequency shift function
\begin{equation}
	 \pi\, \mathcal{H}_{\rm sing}(2\pi\tilde{q}) = (-1)^m\arcsin[\sqrt{\Rm}\cos(4\pi\tilde{q})]\,.
\end{equation}
Taking the derivative one can see that the maximum value of $\partial \mathcal{H}_{\rm sing}\left(2\pi \tilde{q}\right)/\partial\tilde{q}$ is $4\sqrt{\Rm}$ (halfway between a node and an antinode of the field), so that
\begin{equation}
	G_{\rm sing}^{\rm max} =  \frac{FSR}{\lambda}\,4\sqrt{\Rm}\,.
	\label{eq:Gsingmax}
\end{equation}
In order to study the enhancement of the coupling (and the associated optical interference effect)
due to the presence of the second membrane, we have to compare the maximum derivative of the function $\mathcal{H}\left(2\pi \tilde{q}_1,2\pi \tilde{q}_2\right)$ with respect to $4\sqrt{\Rm}$.
In Figure~\ref{fig:Figure_2} we show the cavity mode frequency shifts, and superimposed  the vector plot of
the corresponding gradient field, which gives the values of the two couplings $G_1$ and $G_2$. It shows that the largest optomechanical coupling is achieved simultaneously by the two membranes, and in this case
$G_1 = -G_2$. At this point the cavity mode frequency is sensitive at first order only to the variation of the distance between the two membranes, $q = q_2-q_1$, and is not sensitive to shifts of the CoM of the two membranes, $Q$.
This implies that the coupling of the CoM is zero,  $G_Q = 0$, while that  of the relative coordinate is   $|G_{q} | = |G_j|$ Ref.~\cite{Li:2016aa}.  In this case, in order to determine the gain factor we apply the same argument of Sec. III of Ref.~\cite{Li:2016aa} from Eq.~(19) to Eq.~(23). Specifically, we find that, for $\ell$ integer
\begin{equation}
	|G_j^{\rm max}| = \frac{\sqrt{\Rm} + (-1)^{\ell}\cos[2\pi(\tilde{q}_1 + \tilde{q}_2)]}{1 - \Rm}\,|G_{\rm sing}^{\rm max}|\,.
	\label{eq:Gmax}
\end{equation}
This means that the maximum coupling for both membranes is achieved when $(\tilde{q}_1 + \tilde{q}_2)$ is an integer number
for even $\ell$, and an half-integer for odd $\ell$
(and this is visible also from the vector plots in Figure~\ref{fig:Figure_2}).
Using this condition Eq.~(\ref{eq:Gmax}) reduces to
\begin{equation} \label{eq:gmax}
	|G_j^{\rm max}| = \frac{1}{1 - \sqrt{\Rm}}\,|G_{\rm sing}^{\rm max}|\,.
\end{equation}
In the case of $\Rm = 0.4$, as in our experiment, the optomechanical coupling may increase up to a factor $\sim2.72$.

\new{As discussed in detail in Ref.~\cite{Li:2016aa} (see also \cite{Xuereb:2012fk,Xuereb:2013ys}), the present treatment based on the assumption
$ k_{\ell}^{(0)} \gg \delta k_{\ell}$, allowing to
express the frequency shift explicitly as a function of the empty cavity solution (see Eq.~(\ref{eq:approxzero})), is valid provided that the reflectivity $\Rm$ is not too close to one. This fact could be guessed from the fact that Eqs.~(\ref{eq:Gmax})-(\ref{eq:gmax}) suggest an unlimited value of the optomechanical coupling when $\Rm \to 1$, which is unphysical. In fact, as numerically shown in Ref.~\cite{Li:2016aa} and could be expected also on physical grounds, when $\Rm \geq R \sim 1$ (that is, the membrane reflectivity becomes equal or larger than the cavity mirror reflectivity), Eq.~(\ref{eq:Gmax}) is no more valid, and the optomechanical coupling saturates to a value corresponding to that of the inner Fabry-Perot membrane cavity with length $q$, $|G_j^{\rm sat}|=ck^{(0)}_{\ell}/q = 2\pi c/(\lambda q)$. As underlined in Ref.~\cite{Li:2016aa}, when $|q/L| \ll 1$ and $\Rm \sim R \sim 1$, this saturation value would still correspond to the strong coupling regime where the single-photon optomechanical coupling is equal or larger than the cavity decay rate, because for aligned membranes with negligible absorption, the cavity decay rate remains identical to the value of the main cavity with length $L$. 
In our experiment with commercially available membranes we are far from the condition $\Rm \geq R \sim 1$, and therefore Eqs.~(\ref{eq:Gmax})-(\ref{eq:gmax}) can be safely used to describe the results.}

%------------------------------------------------------------------
\section{Membrane--sandwich characterization}
\label{sec:membrane-sandwich}

In our experiment we used two different membrane sandwiches. The first is constituted of two low-stress SiN square membranes, with a side of  $\SI{1}{\milli\meter}$, and a thickness of \SI{100}{\nano\meter}.
And the second is made of two high-stress $\mathrm{Si_3N_4}$ square membranes, with a side of  $\SI{1.5}{\milli\meter}$, and a nominal thickness of \SI{100}{\nano\meter}.
In both cases, one of the membranes is glued on a piezo, which allows for a scan of the membrane--cavity length, while the whole membrane--cavity mount is attached to another piezo in order to displace in a controlled way the CoM of the two membranes.
\begin{figure}[!b]
\centering
\includegraphics[width=.535\linewidth]{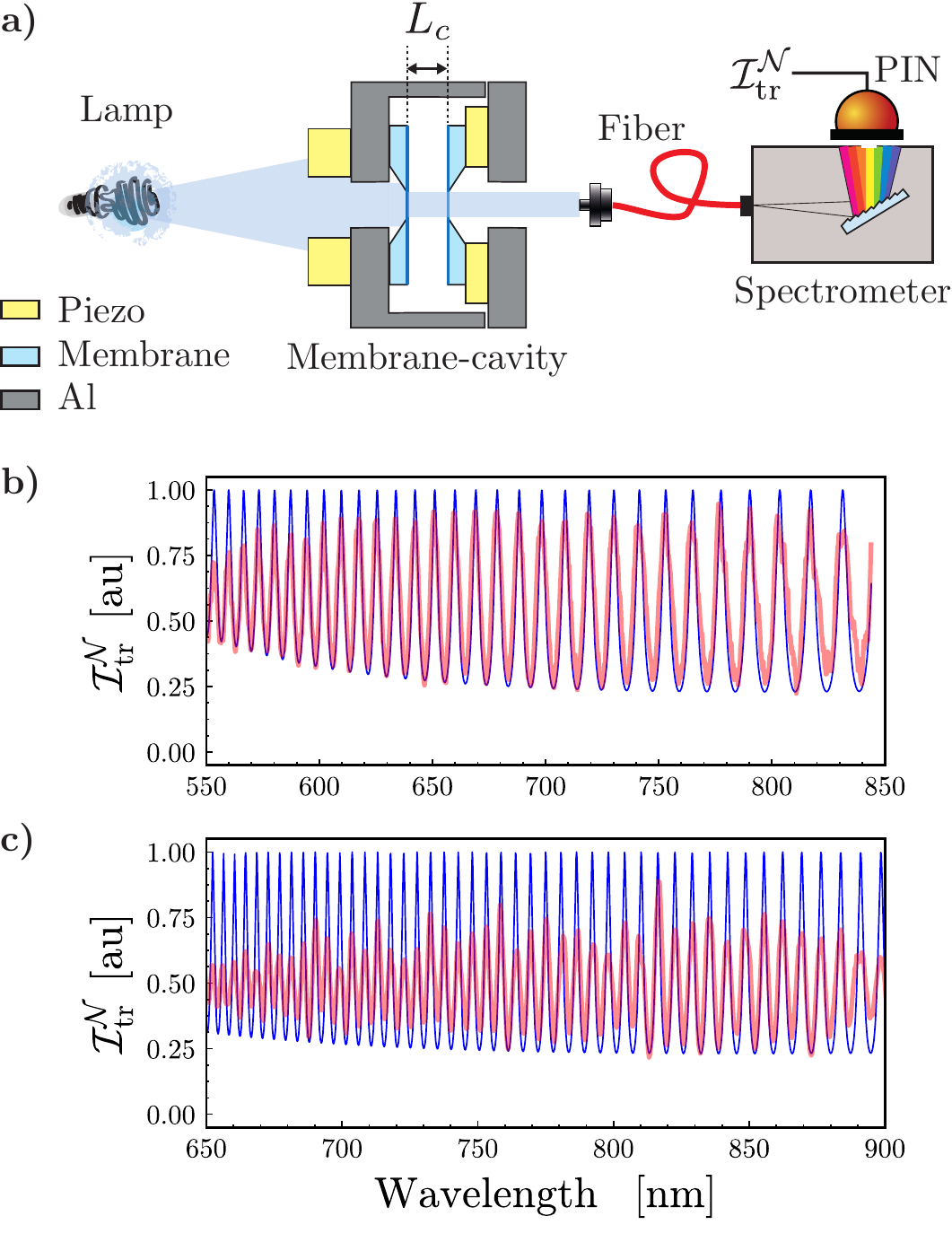}
\caption{
	Cavity--frequency scan.
	a) Experimental setup for cavity frequency--scan. The light of a tungsten lamp transmitted by the membrane sandwich of length $L_c$ at rest, is coupled to a multi-mode optical fiber and collected into a spectrometer for wavelength analysis.
	b) Red line represents the measured light transmitted by the first membrane--cavity, and normalised to the light in the absence of membranes, $\mathcal{I}_\mathrm{tr}^\mathcal{N}$. Blue line is the best--fit obtained with  $L_{\rm c} = \SI{24.008\pm0.004}{\micro\meter}$, and $L_\mathrm{m} = \SI{100.0\pm0.2}{\nano\meter}$.
	c) Red and blue line represent data from the second sandwich and best--fit, respectively. The best--fit provides $L_{\rm c} = \SI{53.571\pm0.009}{\micro\meter}$, and $L_\mathrm{m} = \SI{106\pm1}{\nano\meter}$.}
\label{fig:Figure_3}
\end{figure}

\subsection{Optical properties}

Here we report on the characterization of the two--membrane sandwiches in terms of reflectivity $\Rm $ and cavity length $L_c$, which we have performed before inserting them into the optical cavity.
\begin{figure}[!b]
\centering
\includegraphics[width=.7\linewidth]{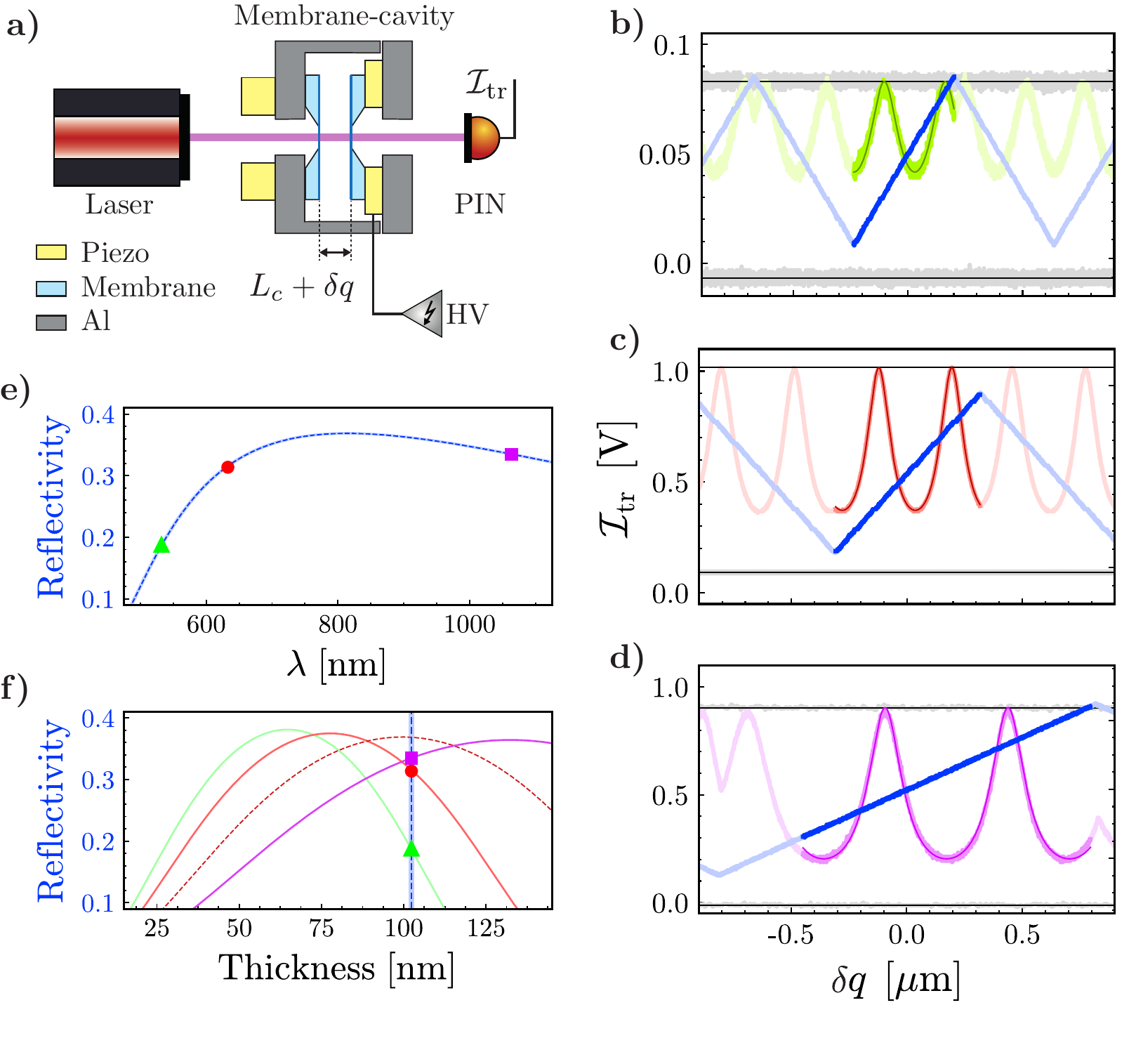}
\caption{
	Cavity--time scan.
	a) Experimental setup for cavity time--scan. A PIN photodiode detects the light transmitted by the membrane--cavity while the membrane distance is scanned by means of a high voltage (HV) applied to a piezo.
	Light transmitted by the membrane--cavity for three different wavelengths, $\SI{532}{\nano\meter}$ (b), $\SI{632.8}{\nano\meter}$ (c), and $\SI{1064}{\nano\meter}$ (d), as a function of the membrane distance $L_c + \delta q$. The best--fit values of the membrane--cavity finesse are $\mathcal{F}_{532}  = \num{1.466\pm0.002}$, $\mathcal{F}_{632.8}  = \num{2.3817\pm0.0007}$, and $\mathcal{F}_{1064}  = \num{3.20\pm0.03}$, which correspond to membrane reflectivities $\mathcal{R}_{\rm m}^{532} = \num{0.2050\pm0.0002}$, $\mathcal{R}_{\rm m}^{632.8} = \num{0.3137\pm0.0001}$, and $\mathcal{R}_{\rm m}^{1064} = \num{0.3345\pm0.0003}$, respectively. Blue line represents the voltage applied to the piezo.
	e) Variation of the reflectivity of the membranes as a function of the wavelength. Green triangle, red circle and  purple square are the measured reflectivity values at $\SI{532}{\nano\meter}$, $\SI{632.8}{\nano\meter}$, and $\SI{1064}{\nano\meter}$, respectively. The best--fit, blue curve, associated to Eq.~(\protect\ref{eq:rm}), provides a value of the membrane thickness of $L_{\rm m} = \SI{102.3\pm0.1}{\nano\meter}$.
	f) Dependence of the reflectivity of a $\mathrm{Si_3N_4}$ membrane on the thickness [Eq.~(\protect\ref{eq:rm})], for three different wavelengths: $\SI{532}{\nano\meter}$, $\SI{632.8}{\nano\meter}$, and $\SI{1064}{\nano\meter}$. Dashed blue line represents the estimated thickness of the measured substrates [$L_{\rm m} = \SI{102.3\pm0.1}{\nano\meter}$].
	}
\label{fig:Figure_4}
\end{figure}
In particular, the membrane--cavity length $L_c$ was determined by illuminating the membrane--sandwich
with a tungsten lamp. The transmitted light
was collected by a multimode fiber, and finally revealed by a spectrometer.
The interference pattern of the normalised transmitted light is shown in Figure~\ref{fig:Figure_3}b) and~\ref{fig:Figure_3}c), for the first and second sandwich, respectively, and compared with a best--fit curve obtained from the expression of the transmitted light
\begin{equation}
	\mathcal{I}_{\rm tr} = \frac{\mathcal{I}_{\rm in}}{1+\left[2\mathcal{F}\sin(\Delta/2)/\pi \right]^2}\, ,
	\label{eq:tr0}
\end{equation}
where $\mathcal{I}_{\rm in}$ is the input light intensity,
$\Delta = 4\pi\,L_c/\lambda$, and $\mathcal{F}$ is the finesse of the membrane--cavity.
From the spectrometer data of the first sandwich, Figure~\ref{fig:Figure_3}b), we obtain a best--fit value for the membrane--cavity length $L_{\rm c} = \SI{24.008\pm0.004}{\micro\meter}$. Moreover, assuming a finesse given by the equation
\begin{equation}
	\mathcal{F} = \frac{\pi}{2} \left[\arcsin\left(\frac{1-\mathcal{R}_{\rm m}}{2\sqrt{\mathcal{R}_{\rm m}}}\right)\right]^{-1}\,,
	\label{eq:fino}
\end{equation}
which holds in the case of equal membrane reflectivity,
and using the values of the index of refraction provided by the manufacturer, we find that the corresponding membrane thickness is $L_\mathrm{m} = \SI{100.0\pm0.2}{\nano\meter}$.
From the data of the second sandwich, Figure~\ref{fig:Figure_3}c), we obtain a membrane--cavity length $L_{\rm c} = \SI{53.571\pm0.009}{\micro\meter}$, and a membrane thickness $L_\mathrm{m} = \SI{106\pm1}{\nano\meter}$,
which is found for the index of refraction of Si$_3$N$_4$ given in Ref.~\cite{Luke:2015aa}.

Although the membrane--cavity length is well estimated by the peak distances in the interference patterns reported in Figure~\ref{fig:Figure_3}b) and~\ref{fig:Figure_3}c), the membrane thickness, and consequently the reflectivity of the membrane, is badly derived by the poor visibility of the curves, measured with an apparatus not optimized for this purpose.
The membrane reflectivity $\Rm$ at specific wavelengths is optimally estimated with a different experiment [see Figure~\ref{fig:Figure_4}a)] exploiting again Eq.~(\ref{eq:tr0})
and (\ref{eq:fino}), but now collecting on a photodiode the light of a laser transmitted through the membrane--cavity while scanning the cavity length $q = L_c + \delta q$, such that, in this case, we use $\Delta = 4\pi\,q/\lambda$ in Eq.~(\ref{eq:tr0}). For the first sandwich we use a $\SI{1064}{\nano\meter}$ laser, and
the best--fit provides a value of the finesse $\mathcal{F}  = \num{3.26\pm0.02}$, yielding a corresponding
value for the reflectivity $\mathcal{R}_{\rm m} = \num{0.408\pm0.002}$.
Such a result is consistent with a membrane thickness of $L_{\rm m} = \SI{104\pm1}{\nano\meter}$, assuming an index of refraction $n= 2.17$. Those values are in accordance with the ones provided by the manufacturer.
For the second sandwich we used three different wavelengths, $\SI{532}{\nano\meter}$, $\SI{632.8}{\nano\meter}$ and $\SI{1064}{\nano\meter}$, and the corresponding results, obtained
while scanning the cavity length, are
shown in Figure~\ref{fig:Figure_4}b)--d). The best--fit of Eq.~(\ref{eq:fino}) provides a value of the finesse and of the corresponding reflectivity for each wavelength. They are given by
$\mathcal{F}_{532}  = \num{1.466\pm0.002}$, $\mathcal{F}_{632.8}  = \num{2.3817\pm0.0007}$, and $\mathcal{F}_{1064}  = \num{3.20\pm0.03}$ with corresponding reflectivity $\mathcal{R}_{\rm m}^{532} = \num{0.2050\pm0.0002}$, $\mathcal{R}_{\rm m}^{632.8} = \num{0.3137\pm0.0001}$, and $\mathcal{R}_{\rm m}^{1064} = \num{0.3345\pm0.0003}$, respectively.
In order to estimate the thickness of the membranes these values were fitted according to the relation in Eq.
(\ref{eq:rm}) [see Figure~\ref{fig:Figure_4}e)]. As shown in Figure~\ref{fig:Figure_4}f), we obtain a membrane thickness of $L_{\rm m} = \SI{102.3\pm0.1}{\nano\meter}$. This result is estimated by using the values of
the refractive index at the three wavelengths reported in~\cite{Luke:2015aa}, which are in accordance with the ones provided by the manufacturer.

\subsection{Mechanical properties}

\begin{figure}[!b]
\centering
\includegraphics[width=.45\linewidth]{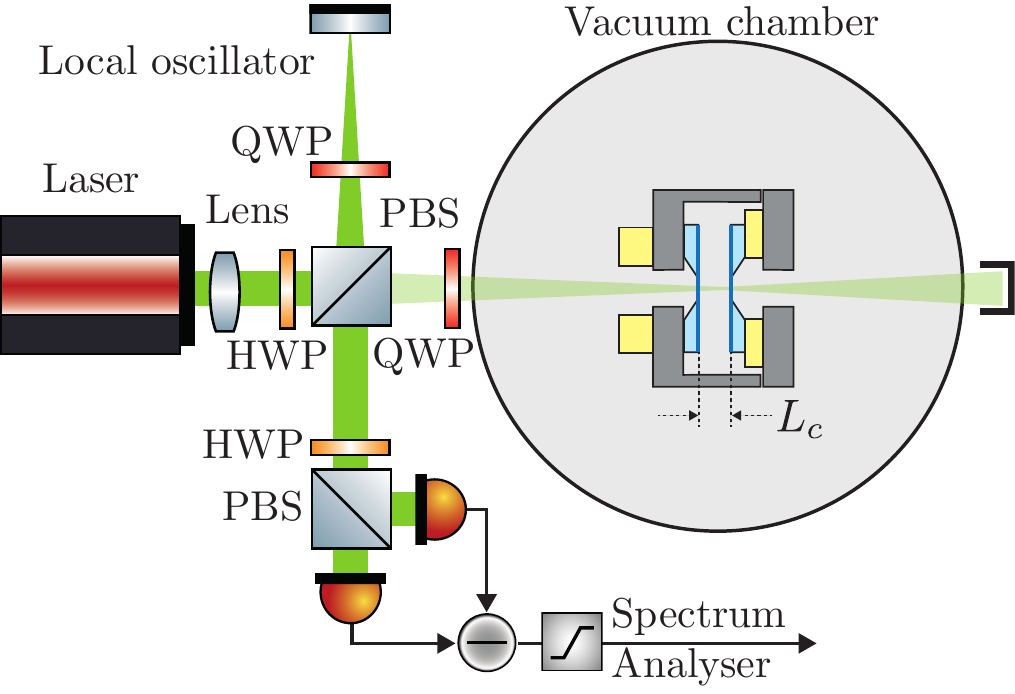}
\caption{
	Experimental setup for characterizing the mechanical properties of the two membranes constituting the membrane--cavity. A $\SI{532}{\nano\meter}$ laser is sent into a polarization--multiplexed  Michelson interferometer. Thermal voltage noise of the two--membrane cavity is revealed by homodyne detection of the reflected light. HWP denotes a half--waveplate, QWP a quarter--waveplate, and PBS a polarizing beam--splitter.
}
\label{fig:Figure_9}
\end{figure}

\begin{figure}[!t]
	\centering
\includegraphics[width=.55\linewidth]{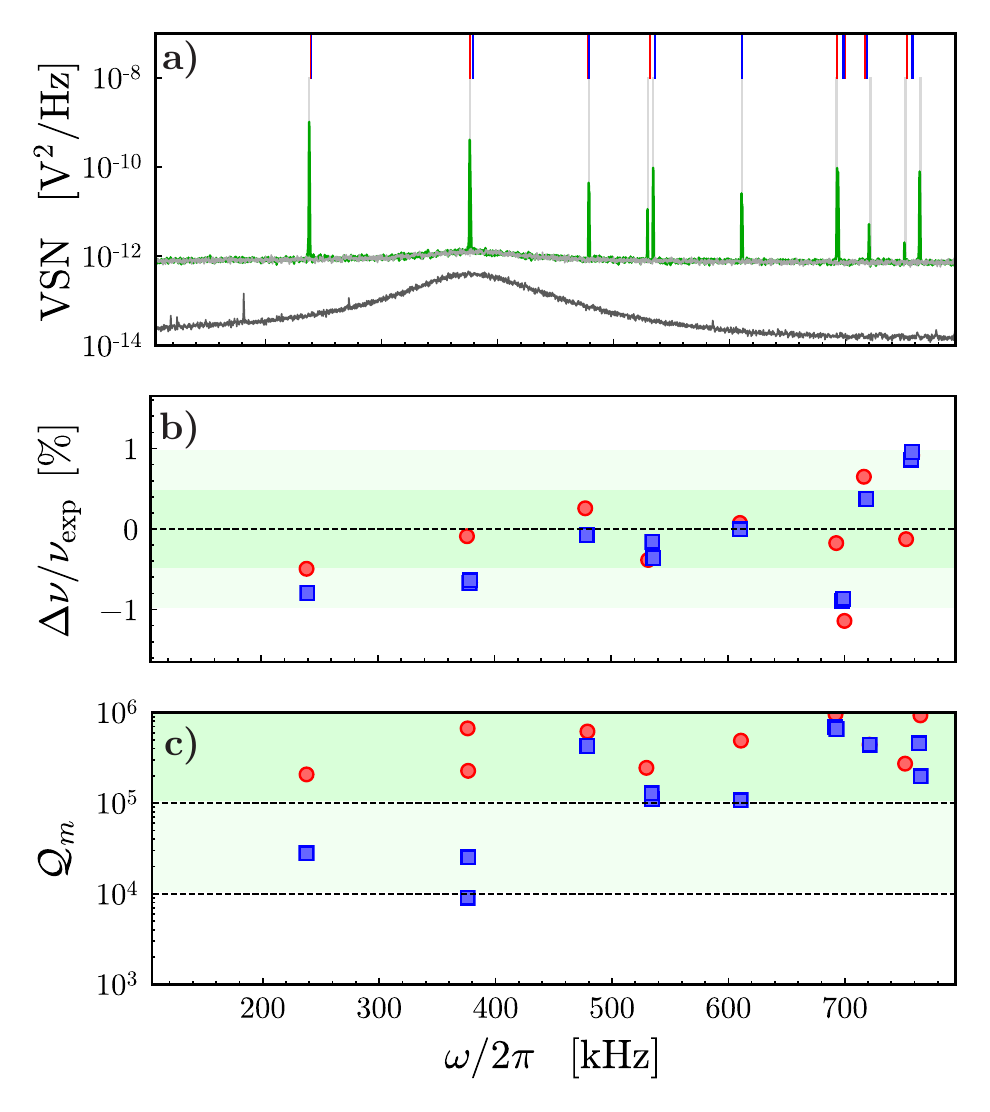}
	\caption{
\new{
	Thermal noise measurement of the mechanical modes of the two membranes in a Michelson interferometer.
	a) Thermal voltage noise (VSN) (green curve) with the experimental mechanical resonance peaks highlighted by vertical light--grey lines; red and blue top lines indicate the mechanical frequencies of rectangular membranes with nominal values for the stress $\sigma = \SI{.825}{\giga\pascal}$ and density $\rho = \SI{3100}{\kilo\gram\per\meter^3}$, and best best--fit parameters for the side lengths
$L_{x}^{(1)} = \SI{1.519 \pm 0.006}{\milli\meter}$, $L_{y}^{(1)} = \SI{ 1.536 \pm0.006}{\milli\meter}$, and $L_{x}^{(2)} = \SI{1.522\pm 0.006}{\milli\meter}$, $L_{y}^{(2)} = \SI{1.525 \pm0.006}{\milli\meter}$, respectively.
The grey curve is the shot noise, while the black curve the electronic noise.
	b) Relative difference between experimental and fitted mechanical frequencies for the two membranes.
	c) Quality factor $\mathcal{Q}_m$ of each mechanical mode.
	}
	}
	\label{fig:Figure_10}
\end{figure}

Here we present a study of the mechanical properties of the second membrane-sandwich by using a $\SI{532}{\nano\meter}$ laser in a Michelson interferometer, as shown in Figure~\ref{fig:Figure_9}~\cite{Bawaj:2015aa}
(this kind of study is not possible with the first sandwich due to the poor quality of the mechanical modes). In Figure~\ref{fig:Figure_10} we show the thermal voltage noise (VSN) of the two--membranes cavity revealed by homodyne detection of the reflected light, the quality factor $\mathcal{Q}_m$ of the mechanical modes, and the relative difference between experimental and fitted mechanical frequencies. The membranes are very similar and show a set of very close resonance peaks.
As shown in Figure~\ref{fig:Figure_10}b), we reproduced the mechanical resonance frequencies of both membranes with an error smaller than $1\%$ assuming rectangular membranes and the nominal values provided by the manufacturer for the stress, $\sigma = \SI{.825}{\giga\pascal}$, and for the density $\rho = \SI{3100}{\kilo\gram\per\meter^3}$, and taking the side lengths as fitting parameters. Best--fit values are
$L_{x}^{(1)} = \SI{1.519 \pm 0.006}{\milli\meter}$, $L_{y}^{(1)} = \SI{ 1.536 \pm0.006}{\milli\meter}$, and $L_{x}^{(2)} = \SI{1.522\pm 0.006}{\milli\meter}$, $L_{y}^{(2)} = \SI{1.525 \pm0.006}{\milli\meter}$. Figure~\ref{fig:Figure_10}c) shows that the mechanical quality factor changes significantly between the modes and that one membrane tends to have lower $\mathcal{Q}_m$ values. We attribute these scattered values to the effect of clamping which strongly depends upon the shape of the vibrational mode and may be different on the two membranes with the current mounting.

%------------------------------------------------------------------

\section{Estimation of the optomechanical coupling strength}

\label{sec:optical properties}

\begin{figure}[!t]
\centering
\includegraphics[width=.5\linewidth]{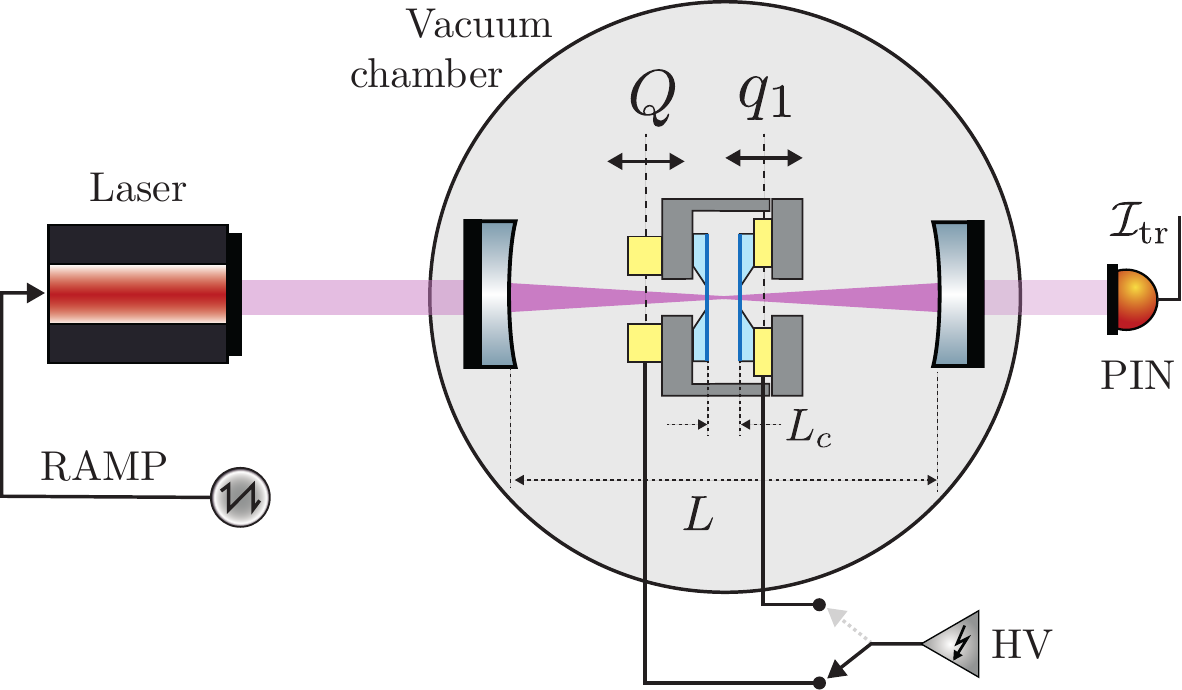}
\caption{
	Experimental setup for the measurements reported in Figures~\ref{fig:Figure_6}, and~\ref{fig:Figure_7}.
	The light of a laser at $\SI{1064}{\nano\meter}$ wavelength transmitted by an optical cavity of length $L = \SI{90}{\milli\meter}$  containing the membrane sandwich of thickness $L_m = \SI{104}{\nano\meter}$, and distance $L_c = \SI{24}{\micro\meter}$ at rest, is revealed by a PIN photodiode ($\mathcal{I}_\mathrm{tr}^\mathcal{N}$), while the frequency is scanned by applying a ramp signal (RAMP) to the piezo control of the laser. The positions of the two membranes are controlled by applying high-voltage (HV) to the piezos, which move the CoM, $Q$, and the cavity length, $q_1$.
}
\label{fig:Figure_5}
\end{figure}

In order to estimate the strength of the optomechanical coupling achievable with our system we have inserted the first sandwich (the one made with the SiN membranes) in a \SI{90}{\milli\meter}--length optical cavity~\cite{Rossi:2017aa,Kralj:2017aa}, and the optomechanical system was located in a vacuum chamber evacuated to $\SI{5e-7}{\milli\bar}$ (see Figure~\ref{fig:Figure_5}).
\begin{figure}[!b]
\centering
\includegraphics[width=.7\linewidth]{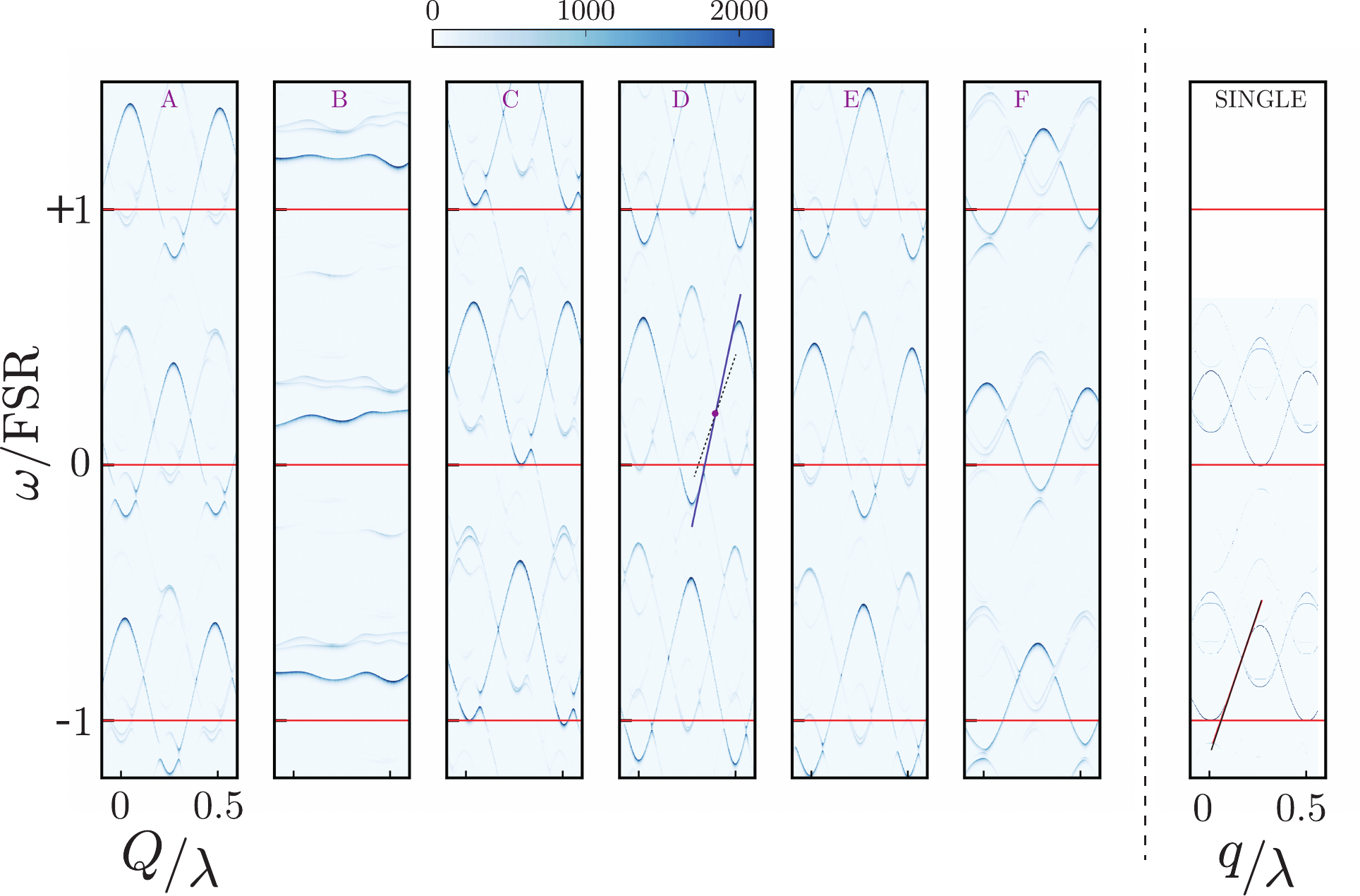}
\caption{Mode frequency shift normalized to the FSR, as a function of the CoM, $Q$, normalized to the wavelength, for different values of the distance $q=q_2-q_1$ as indicated by the lines A--F in Figure~\ref{fig:Figure_2}. Panel D shows the position of the highest achievable coupling $G_Q^{\rm max} \simeq 2\pi\times\SI{5.67}{\mega\hertz\per\nano\meter}$ indicated by the solid blue line.
	For comparison the single--membrane result is added as a dotted black line, which represents the maximum achievable coupling $G_{\rm sing}^{\rm max} \simeq 2\pi\times\SI{3.47}{\mega\hertz\per\nano\meter}$, shown in the panel on the right.}
\label{fig:Figure_6}
\end{figure}
\begin{figure}[!t]
\centering
\includegraphics[width=.7\linewidth]{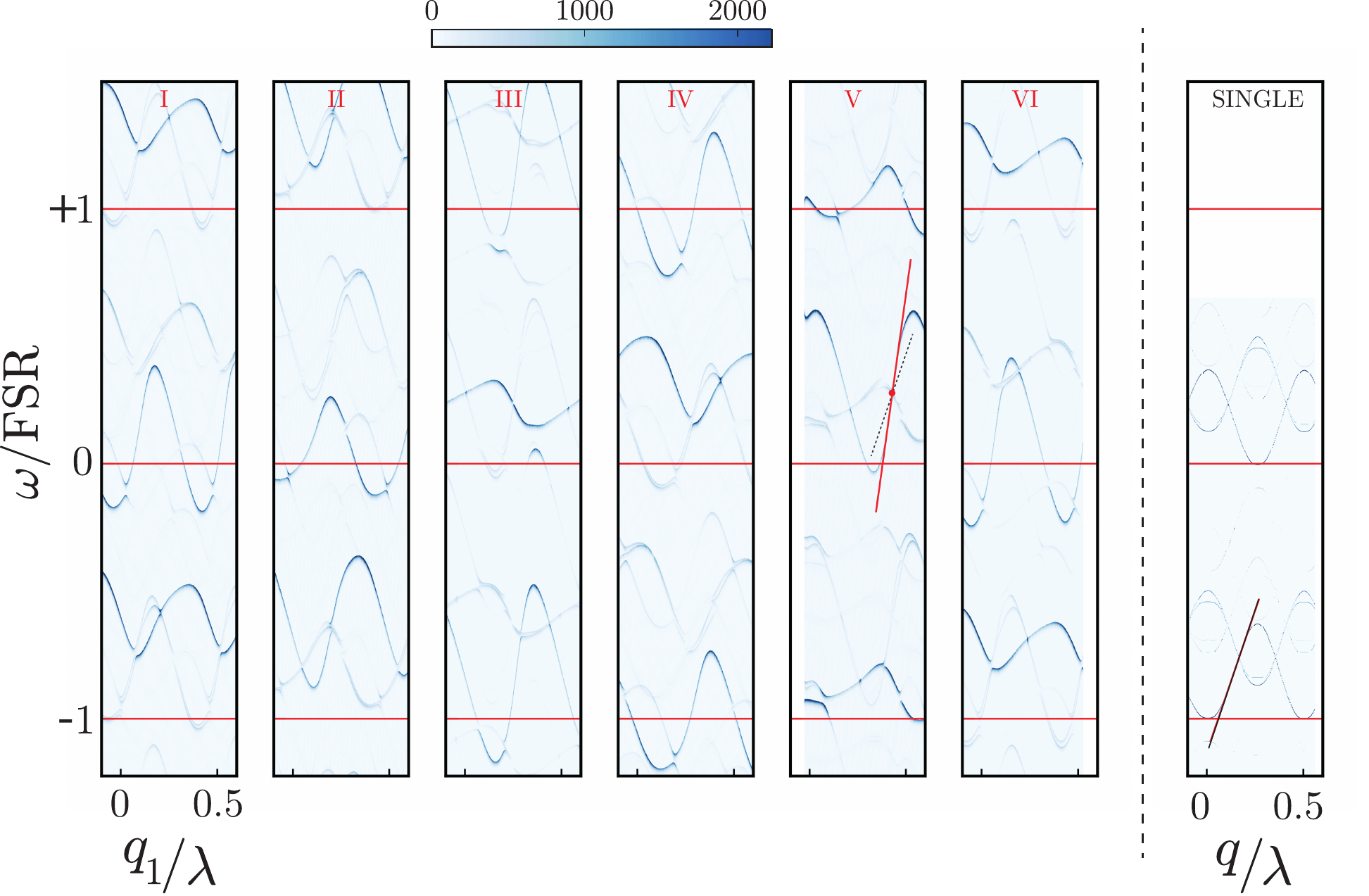}
\caption{Mode frequency shift normalized to the FSR, as a function of the membrane position $q_1$, normalized to the wavelength, for different values of the position $q_2$, as indicated by the lines I--VI in Figure~\ref{fig:Figure_2}. Panel V shows the positions for the highest coupling $G_1^{\rm max} \simeq 2\pi\times\SI{8.59}{\mega\hertz\per\nano\meter}$.
	For comparison the single--membrane result is added as a dotted black line, as in Figure~\ref{fig:Figure_6}.}
\label{fig:Figure_7}
\end{figure}

Our aim is to compare the frequency shift of the resulting cavity modes in the presence of the two-membrane system, with the one corresponding to the case with a single membrane inside. We note that the results for a single membrane are obtained using a membrane different form the ones of the sandwich, namely a highly stressed SiN circular membrane, with a diameter of \SI{1.2}{\milli\meter}, and a thickness of \SI{97} {\nano\meter}~\cite{Serra:2016aa,Rossi:2017aa,Kralj:2017aa}.
However, the fact that the membranes have similar size and are made of the same material, makes the comparison that we report hereafter meaningful.

The spectra of the cavity modes reported in Figures~\ref{fig:Figure_6} and~\ref{fig:Figure_7} are obtained by detecting the light of a laser at $\SI{1064}{\nano\meter}$ transmitted by the cavity while scanning the laser frequency for different positions of the membrane(s).
The last panel on the right of Figure~\ref{fig:Figure_6} is equal to the last of Figure~\ref{fig:Figure_7}
and they report the results of the single membrane case. The slope of the corresponding black lines represents the maximum achievable single--membrane optomechanical coupling strength $G_{\rm sing}^{\rm max} \simeq 2\pi\times\SI{3.47}{\mega\hertz\per\nano\meter}$.
The other panels show the results with two membranes.
In this case there are two degrees of freedom that can be varied, that is, the positions of the two membranes, $q_1$ and $q_2$. Due to the design of our membrane--cavity, we can scan either the CoM, $Q$, for different values of the membrane distance $q = q_2 - q_1$, or $q_1$ for different positions of $q_2$.
In Figure~\ref{fig:Figure_6} are reported the spectra obtained by scanning the CoM, $Q$, for different values of the membrane distance $q$, as indicated by the lines A--F in Figure~\ref{fig:Figure_2}.
The blue line  on panel D corresponds to the blue circle in Figure~\ref{fig:Figure_2}, and it indicates the highest
coupling $G_Q^{\rm max} \simeq 2\pi\times\SI{5.67}{\mega\hertz\per\nano\meter}$ achieved in this case.
It corresponds to an increase in the optomechanical coupling strength of a factor $\sim\num{1.63}$ with respect to the single--membrane case.
In Figure~\ref{fig:Figure_7} we report the spectra obtained by scanning the position $q_1$ for different position $q_2$, as indicated by the lines I--VI in Figure~\ref{fig:Figure_2}. The red line on panel V  corresponds to the red circle in Figure~\ref{fig:Figure_2}, and indicates the highest achieved coupling $G_1^{\rm max} \simeq 2\pi\times\SI{8.59}{\mega\hertz\per\nano\meter}$.  In this case the optomechanical coupling strength increases by a factor $\sim\num{2.47}$, which is $\SI{9}{\percent}$ lower than the expected one, given by Eq.~(\ref{eq:gmax}). Such a discrepancy might be attributed to an imperfect alignment of the two membranes.

%------------------------------------------------------------------

\section{Cavity finesse in the presence of the membrane-sandwich}
\label{Sec.CavityFiness}

\begin{figure}[!b]
\centering
\includegraphics[width=.6\linewidth]{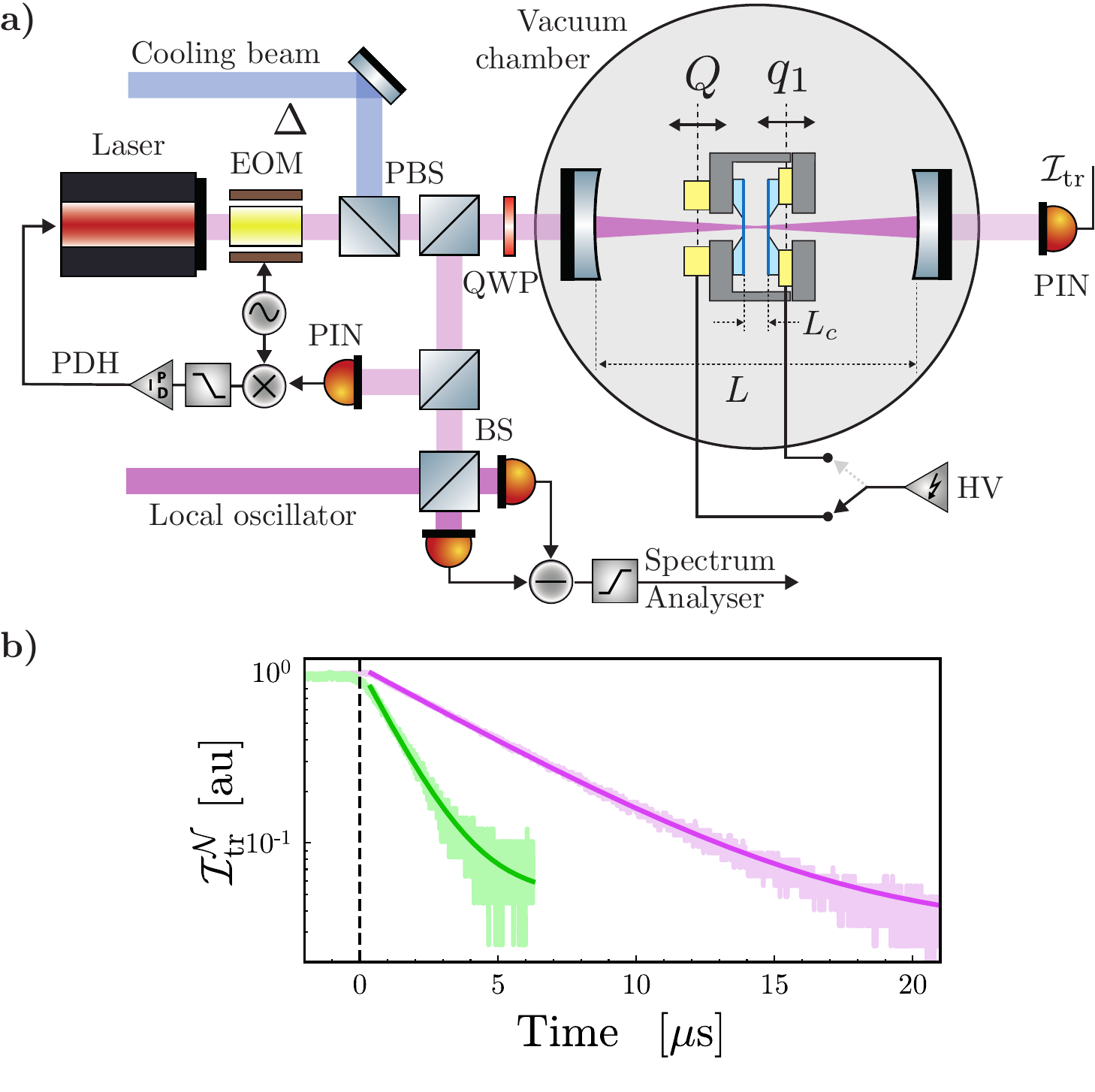}
\caption{
	a) Experimental setup for studying cavity optomechanics with a two-membrane setup within a cavity.
	A laser probe beam, frequency modulated by an electro-optical modulator (EOM), impinges on the optical cavity. The reflected beam is split: one component is detected, demodulated and low-pass amplified for generating the Pound--Drever--Hall (PDH) error signal able to lock the laser to the cavity; the second component is analyzed by homodyne detection in order to detect the mechanical motion. A further beam, the cooling beam, detuned by $\Delta$ from the cavity resonance, is turned on for engineering the optomechanical interaction, and in particular realize laser cooling of the mechanical modes. HWP denotes a half--waveplate, QWP a quarter--waveplate, BS a beam--splitter, and PBS a polarizing beam--splitter.
	b) Cavity ringdown measurement for the evaluation of the cavity finesse. Light--violet data is the normalized transmitted intensity, $\mathcal{I}_\mathrm{tr}^\mathcal{N}$, through the empty optical cavity; the solid violet line represents the best--fit with decay time $\tau_0 = \SI{4.790\pm0.002}{\micro\second}$, which corresponds to an empty cavity finesse $\mathcal{F}_0 = \pi\tau_0\,c/L = \si{50125\pm25}$. Light green data refer to the case with the membrane--sandwich placed within the optical cavity; the solid green line is the best--fit with decay time $\tau = \SI{1.365\pm0.001}{\micro\second}$, corresponding to a finesse $\mathcal{F} = \si{14287\pm13}$.
}
\label{fig:Figure_11}
\end{figure}

In the last set of experiments we placed the second membrane sandwich (the one made of $\mathrm{Si_3N_4}$ membranes) in the same optical cavity of Figure~\ref{fig:Figure_5} [see also Fig.~\ref{fig:Figure_11}a)].
Here we report on the analysis of the effects of the membranes on the cavity finesse. The
finesse of the optical cavity, with and without the membrane sandwich, is determined by means of the ring--down technique, fitting the decay of the normalized transmitted intensity, $\mathcal{I}_\mathrm{tr}^\mathcal{N}$, after the laser at $\SI{1064}{\nano\meter}$ is rapidly turned off. In Figure~\ref{fig:Figure_11}b) we show the ring--down results obtained for the empty cavity, and with the membrane-sandwich placed within the optical cavity.
For the former case, the best--fit decay time is $\tau_0 = \SI{4.790\pm0.002}{\micro\second}$, which corresponds to an empty cavity finesse $\mathcal{F}_0 = \pi\tau_0\,c/L = \si{50125\pm25}$~\cite{Rossi:2017aa}, while for the latter, $\tau = \SI{1.365\pm0.001}{\micro\second}$, corresponding to a cavity finesse $\mathcal{F} = \si{14287\pm13}$.
Such finesse corresponds to a cavity intensity decay rate $\kappa = \tau^{-1} = \mathrm{FSR} / \mathcal{F} \sim 2\pi\times\SI{117}{\kilo\hertz}$, with $\mathrm{FSR} \sim 2\pi\times\SI{1.67}{\giga\hertz}$.
The observed reduction of finesse in the presence of the membrane-sandwich is much more significant than the one occurring in the case of a single membrane \cite{Wilson:2009a,Serra:2016aa} and it can be ascribed to the imperfect alignment of the two membranes~\cite{Nair:2017ab}. This misalignment is responsible for an effective cavity loss $1/\delta\mathcal{F} =  1/\mathcal{F} - 1/\mathcal{F}_0 \simeq (\sqrt{F_{m}}\,\theta_\mathrm{wdg}/\theta_\mathrm{dif})^2 / 2\pi \simeq \SI{50}{ppm}$. Assuming a coefficient of finesse $F_m = 4\mathcal{R}_\mathrm{m}/(1 - \mathcal{R}_\mathrm{m})^2 \simeq 3$, and a diffraction angle of the gaussian beam $\theta_\mathrm{dif} = \lambda/\pi w_0 \simeq \SI{3}{\milli\radian}$, with $w_0 \simeq \SI{112}{\micro\meter}$ the beam waist of the cavity of our experiment, the misalignment angle $\theta_\mathrm{wdg}$ between the two non-parallel membranes can then be estimated to be $\theta_\mathrm{wdg} \sim \SI{30}{\micro\radian}$.
\new{The membrane alignment could be improved either by using pairs of membranes assembled parallel to each other by means of spacers deposited on one of the chip, as implemented for example in the experiment of Ref.~\cite{Nair:2017ab}, or by replacing the single piezo, used for the scan of the membrane--cavity, with tilt stages with piezo control, which would allow for scanning as well as alignment of  the membrane--cavity.}

%s, or (2) on top adding for example a dedicated piezocontrol and employing a tunable laser, as implemented for example in the experiment of Ref.~\cite{Nair:2017ab}.}

\section{Tunable optomechanical coupling and laser cooling of the two membranes}

Using the same setup of Sec.~\ref{Sec.CavityFiness}, we finally studied the optomechanical properties of the system.
\begin{figure}[!b]
	\centering
\includegraphics[width=.45\linewidth]{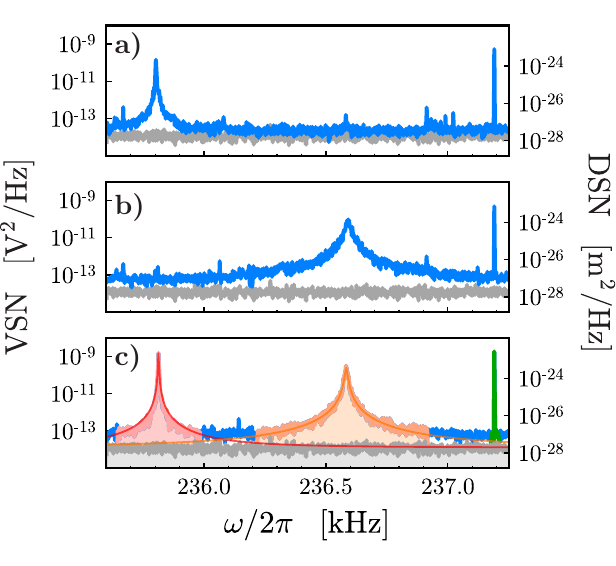}
	\caption{
	Thermal voltage (VSN) and displacement (DSN) spectral noise of the membrane sandwich obtained by homodyne detection of the light reflected by the optical cavity.
	a) Only the membrane with lower frequency fundamental mode is coupled to the optical cavity.
	b) Only the membrane with higher frequency fundamental mode is coupled to the optical cavity.
	c) The fundamental modes of both membranes are coupled to the optical cavity. The green feature on the right indicates the beat note added for calibration. For the left red mode we determine: $\omega_{m1} = 2\pi\times\SI{235.810}{\kilo\hertz}$, $\gamma_{m1} = 2\pi\times\SI{1.64}{\hertz}$, and $g_{01} = 2\pi\times\SI{0.30}{\hertz}$; and For the right blue mode: $\omega_{m2} = 2\pi\times\SI{236.580}{\kilo\hertz}$, $\gamma_{m2} = 2\pi\times\SI{9.37}{\hertz}$, and $g_{02} = 2\pi\times\SI{0.28}{\hertz}$.
	}
	\label{fig:Figure_12}
\end{figure}
First we show that the optomechanical interaction of the driven cavity mode with each membrane of the sandwich can be controlled and tuned by shifting their position along the cavity axis with the piezo controllers. The probe beam was locked to the optical cavity by means of a Pound--Drever--Hall (PDH) technique and the thermal voltage spectral noise (VSN) of the two--membranes cavity is measured by homodyne detection of the light reflected by the optical cavity [see Figure~\ref{fig:Figure_11}a)]. The detected thermal voltage spectral noise is shown in Figure~\ref{fig:Figure_12}, which clearly manifests the possibility to turn on and off the optomechanical interaction in a controlled manner by changing the position of each membrane [see Figure~\ref{fig:Figure_12}a) and Figure~\ref{fig:Figure_12}b)] where only one of the two membranes is positioned in a place in which it interacts with the cavity light). In Figure~\ref{fig:Figure_12}c)
both membranes are instead coupled to the optical cavity. For the lower frequency mode on the left (red) we measured $\omega_{m1} = 2\pi\times\SI{235.810}{\kilo\hertz}$, $\gamma_{m1} = 2\pi\times\SI{1.64}{\hertz}$, while for the mode on the right (orange) we measured $\omega_{m2} = 2\pi\times\SI{236.580}{\kilo\hertz}$, $\gamma_{m2} = 2\pi\times\SI{9.37}{\hertz}$.
\begin{figure}[!b]
	\centering
	\includegraphics[width=.65\linewidth]{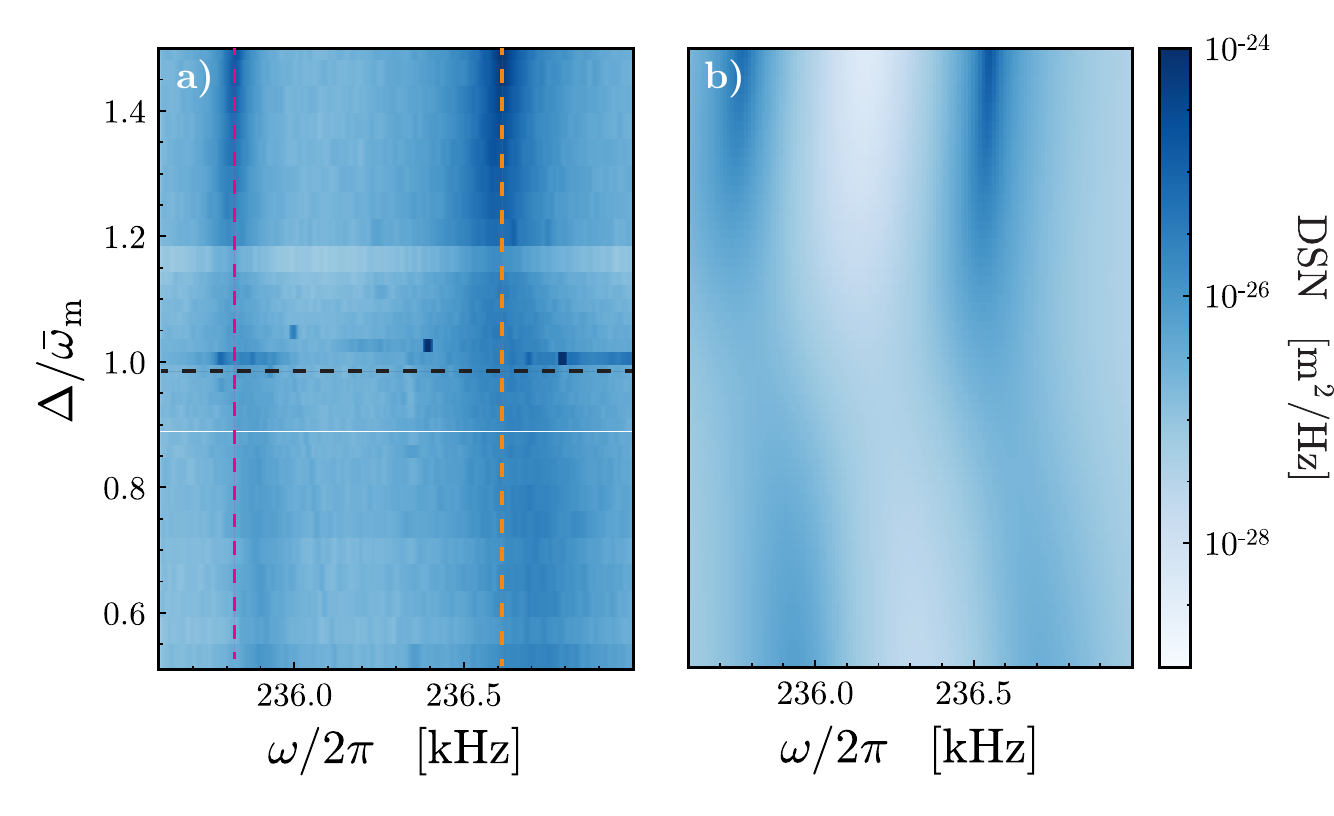}
	\caption{
	%{\bf NUOVA FIGURA}
	Laser cooling of the two membranes at low power.
	a) Measured displacement spectral noise (DSN) as a function of the detuning $\Delta$ normalized to the mean mechanical frequency $\bar{\omega}_m = (\omega_{m1} + \omega_{m2})/2$, for a cooling input power $P_C = \SI{130}{\micro\watt}$,  $\kappa = 2\pi\times\SI{83}{\kilo\hertz}$, and $g_0$ as in Figure~\ref{fig:Figure_12}. The red and orange dashed lines indicate the mechanical frequencies with no cooling.
	b) Theoretical prediction with parameters given in Figure~\ref{fig:Figure_12}.
	}
	\label{fig:Figure_13}
\end{figure}
\begin{figure}[!t]
	\centering
	\includegraphics[width=.65\linewidth]{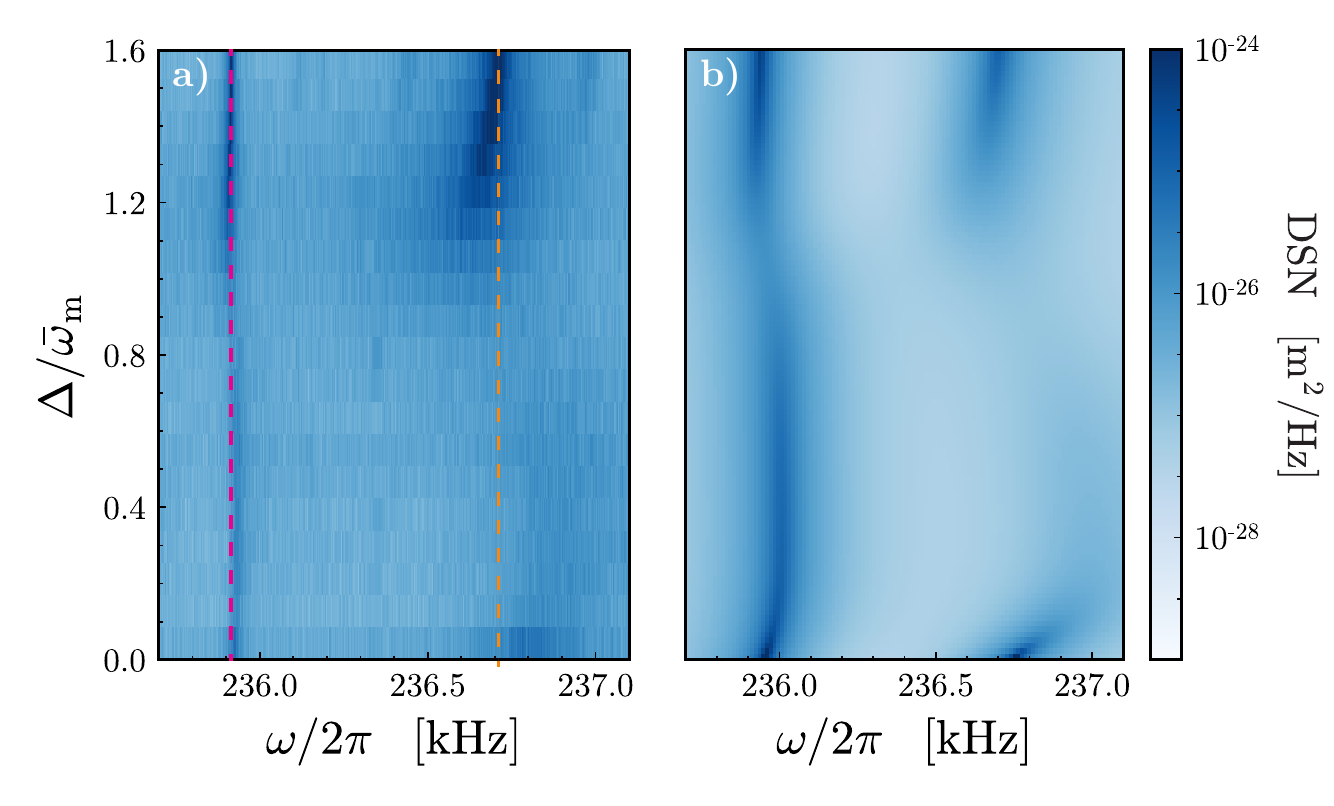}
	\caption{
	Laser cooling of the two membranes at high power.
	a) Measured displacement spectral noise (DSN) as a function of the detuning $\Delta$ normalized to the mean mechanical frequency $\bar{\omega}_m = (\omega_{m1} + \omega_{m2})/2$, for a cooling input power $P_C = \SI{380}{\micro\watt}$, and $\kappa = 2\pi\times\SI{83}{\kilo\hertz}$. The red and orange dashed lines indicate the mechanical frequencies with no cooling.
	b) Theoretical prediction with the following parameters: $\omega_{m1} = 2\pi\times\SI{235.950}{\kilo\hertz}$, $\gamma_{m1} = 2\pi\times\SI{1.64}{\hertz}$, and $g_{01} = 2\pi\times\SI{0.12}{\hertz}$; and For the right blue mode: $\omega_{m2} = 2\pi\times\SI{236.750}{\kilo\hertz}$, $\gamma_{m2} = 2\pi\times\SI{9.37}{\hertz}$, and $g_{02} = 2\pi\times\SI{0.22}{\hertz}$. Note the less effective optomechanical cooling on the left mode due to lower optomechanical coupling, and also the frequency shift in the moderate resolved--side--band limit.
	}
	\label{fig:Figure_14}
\end{figure}
\begin{figure}[!b]
	\centering
	\includegraphics[width=.65\linewidth]{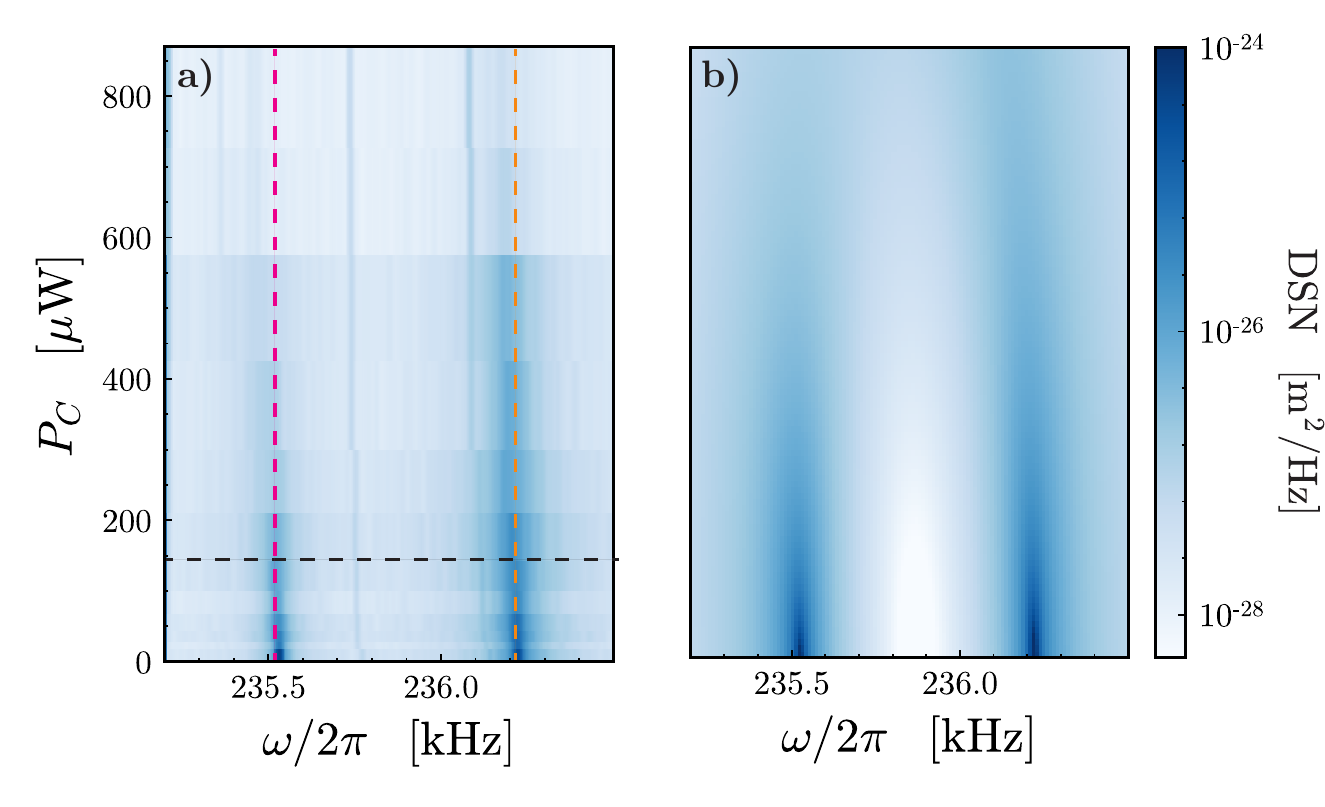}
	\caption{
	Laser cooling of the two membranes at constant detuning.
	a) Measured displacement spectral noise (DSN) as a function of the cooling beam power $P_C$. The red and orange dashed lines indicate the mechanical frequencies with no cooling.
	b) Theoretical prediction for a detuning $\Delta \sim \bar{\omega}_m$ with the same experimental parameters as in Figure~\ref{fig:Figure_12}.
	}
	\label{fig:Figure_15}
\end{figure}
Such results are consistent with the measurements obtained with the interferometer (see Figure~\ref{fig:Figure_10}). In fact, we used a probe beam with very low power, and as resonant as possible with a cavity mode
in order to avoid any optomechanical effect, such as cooling or optical spring effect,
taking into account that $\kappa \sim \bar{\omega}_m/2$ with $\bar{\omega}_m = (\omega_{m1} + \omega_{m2})/2$.
The corresponding measured single--photon optomechanical coupling rates  $g_{0j} = G_j x_{j}^{\rm zpf}\Theta_j$,
where $x_{j}^{\rm zpf} = [\hbar/2m_j\omega_{\rm m}^{(j)}]^{1/2}$ is the zero point position fluctuations of the $j$-th mechanical mode, and $\Theta_j$ is the dimensionless transverse overlap between the $j$-th mechanical mode and the optical cavity mode,~\cite{Gorodetsky:2010uq} are $g_{01} = 2\pi\times\SI{0.30}{\hertz}$ and $g_{02} = 2\pi\times\SI{0.28}{\hertz}$.
These values are comparable to those achieved in a similar setup with a single membrane~\cite{Rossi:2017aa,Rossi:2018aa} because the two membranes were placed out of the region in the $q_1, q_2$ plane where the optomechanical coupling is enhanced due to interference (see Figure~\ref{fig:Figure_2}). Within this region the system was not stable enough and we did not carry out cavity optomechanics experiments.

Finally we show that we can engineer the optomechanical interaction of both membranes with the optical mode by turning on an additional ``cooling'' beam with a variable detuning $\Delta$ with respect the cavity resonance. Here we focus on the case of red-detuned driving which resonantly enhances the beam-splitter interaction between the cavity mode and the mechanical modes and allows to cool the latter. We observe the simultaneous cooling \cite{Genes:2008c} of the fundamental modes of the two distinct membranes. In Figures~\ref{fig:Figure_13} and~\ref{fig:Figure_14} we report the measured displacement spectral noise (DSN) (left panels) as a function of the detuning $\Delta$ normalized to the mean mechanical frequency $\bar{\omega}_m$, and compare it with the corresponding theoretical prediction (right panels). In Figure~\ref{fig:Figure_13} we use a lower power of the cooling beam with respect to that used in Figure~\ref{fig:Figure_14}, but in both cases the agreement is very good. In Figure~\ref{fig:Figure_15} instead we report the DSN as a function of the cooling beam power $P_C$, at a fixed detuning $\Delta \sim \bar{\omega}_m$,

\section{Conclusion}
We studied the optomechanical behaviour of a driven Fabry--P\'erot cavity containing a two-membrane sandwich. From the cavity--mode frequency shift as a function of the membrane positions, we derived a $\sim 2.47$ gain in the
optomechanical coupling strength with respect to the single-membrane case. This is  obtained when the two membranes are positioned to form an inner cavity resonant to the driving field.
We also showed the capability of the system to be tuned on demand, and the simultaneous optical cooling of the fundamental modes of the two distinct membranes.
Such a configuration has the potential to enable cavity optomechanics in the strong single-photon coupling regime~\cite{Xuereb:2013ys,Li:2016aa,Xuereb:2012fk}, as well as to study the nonlinear dynamics and synchronization of two distinct nanomechanical resonators by means of an optical link~\cite{Zhang:2012aa,Bagheri:2013ht,Agrawal:2013fk,Matheny:2014fv,Zhang:2015ad}.

%\section*{Funding Information}

\section*{Acknowledgments}
Authors thank M. Rossi, N. Kralj, A. Borrielli, G. Pandraud, and
E. Serra for the single--membrane spectrum obtained in the alignment procedure of the optomechanical system reported in Refs.~\cite{Rossi:2017aa,Kralj:2017aa,Rossi:2018aa}.
Authors thank also R. Gunnella for the use of the spectrometer, and R. Tossici for optical fiber patches. P. Piergentili acknowledges support from the European Union's Horizon 2020 Programme for Research and Innovation under grant agreement No. 722923 (Marie Curie ETN - OMT).
We also acknowledge the support of the European Union Horizon 2020 Programme for Research and Innovation through the Project No. 732894 (FET Proactive HOT).

\section*{References}
% Bibliography
\bibliographystyle{iopart-num}
\bibliography{Paper_Biblio}

\end{document}